\RequirePackage{ifluatex}
\documentclass
[twocolumn,aps,prd,amsmath,amssymb,nofootinbib,floatfix]
{revtex4-1}

\usepackage{CJK}                     
\usepackage[dvips]{graphicx}         
\usepackage{physics}                 
\usepackage{mathptmx}                
\usepackage{dcolumn}                 
\usepackage{multirow}                
\usepackage{xcolor}                  %
\usepackage{hyperref}
\hypersetup{
 colorlinks=true,     
 linkcolor=blue,      
 citecolor=cyan,      
 filecolor=magenta,   
 urlcolor=cyan         
}
\graphicspath{
{Figures/}
}

\def\be{\begin{equation}} \def\ee{\end{equation}}
\def\bal#1\eal{\begin{align}#1\end{align}}
\def\bse#1\ese{\begin{subequations}#1\end{subequations}}

\def\eps{\epsilon}
\def\la{\Lambda}
\def\ms{\,M_\odot}
\def\mmax{M_\text{max}}
\def\mtov{M_\text{TOV}}
\def\mev{\;\text{MeV}}
\def\fm3{\,\text{fm}^{-3}}
\def\gc3{\,\text{g/cm}^3}
\def\mgw{M_B^\text{GW170817}}
\def\msim{M_B^\text{sim}}
\def\mdk{M_\text{disk}}

\begin{document}

\title{
Binary neutron star merger simulations with hot microscopic equations of state}

\begin{CJK*}{UTF8}{gbsn}

\author{A. Figura$^{1}$}
\author{Fan Li (李凡)$^2$}
\author{Jia-Jing Lu (陆家靖)$^2$}
\author{G. F. Burgio$^1$}
\author{Zeng-Hua Li (李增花)$^2$}
\author{H.-J. Schulze$^1$}

\affiliation{
\vbox{
$^1$ INFN Sezione di Catania, Dipartimento di Fisica,
Universit\'a di Catania, Via Santa Sofia 64, 95123 Catania, Italy}
\vbox{
$^2$ Institute of Modern Physics,
Key Laboratory of Nuclear Physics and Ion-beam Application (MOE),}
Fudan University, Shanghai 200433, P.R.~China
}

\date{\today}

\begin{abstract}
We perform binary neutron star merger simulations
using a newly derived set of finite-temperature equations of state
in the Brueckner-Hartree-Fock approach.
We point out the important and opposite roles of finite temperature
and rotation for stellar stability
and systematically investigate the gravitational-wave properties,
matter distribution, and ejecta properties
in the postmerger phase
for the different cases.
The validity of several universal relations is also examined
and the most suitable EOSs are identified.
\end{abstract}


\maketitle
\end{CJK*}

\section{Introduction}

After the first detection of gravitational waves (GWs)
from a binary neutron star merger (BNSM) \cite{Abbott2017_etal},
and its related electromagnetic counterparts \cite{Abbott2017b},
the constraints on the equation of state (EOS) governing nuclear matter
in neutron stars (NSs) have been impressively refined
\cite{Margalit2017,Rezzolla2017,Radice2017b,Paschalidis2018a,Burgio2018,Wei2019}.
In particular, GWs from the inspiral phase have already set constraints
on the EOS at zero temperature,
showing consistency with small neutron star radii and tidal deformabilities
\cite{Abbott2018b}.
The future detection of a post-merger GW signal could instead
provide information on the EOS of hot
(with temperatures of typically several tens of MeVs
\cite{Perego2019,Figura2020})
and dense
(a few times nuclear saturation density
$\rho_0 \sim 2.7 \times 10^{14}\gc3$) nuclear matter:
interestingly, such signal could also be indicative of the appearance of
deconfined quark matter
\cite{Bauswein2012a,Rezzolla2016,Bauswein2019,Most2019c}.
Due to the frenetic theoretical activity in this field,
the relevant literature is vast and we refer to recent reviews
\cite{Baiotti:2019sew,Radice2020} for an overview.

The accurate simulation of BNSMs represents in this context
a necessary tool to analyze both the GW signal
and the hydrodynamic properties involved in these phenomena,
and the use of a constraint-fulfilling, realistic EOS
is an essential requirement.
Such an EOS should cover \cite{Rezzolla2018} a typical range of about
$10^4 < \rho < 10^{15}\gc3$ in rest-mass densities and
$0 \le T \lesssim 100\mev$ in temperatures,
and should be computed without postulating beta-equilibrium,
accounting for electron fractions in a typical range
$0 \le Y_e \le 0.65$.

EOSs of this kind are very few and generally employ phenomenological models
rather than ab-initio calculations;
noteworthy cases are Relativistic Mean Field (RMF) models
such as the Shen EOS \cite{Shen11},
the DD2 EOS \cite{Typel2010},
the SFHo EOS \cite{Hempel2012,Steiner2013},
or the BHB$\Lambda \phi$ EOS \cite{Banik2014},
which also accounts for hyperon-hyperon interactions.
We also mention the very commonly used LS220 EOS \cite{Lattimer91},
a model based on nonrelativistic Skyrme interactions,
the Togashi model \cite{Togashi2016},
based on a variational approach,
and the recent Chiral Mean Field (CMF) theory based EOS presented in
Ref.~\cite{Most2019c},
where also the possible deconfinement to quark matter is considered.

Given the restricted number of publicly available and constraint-fulfilling
finite-temperature EOSs,
BNSM simulations are also usually performed using the so-called
``hybrid-EOS'' approach
\cite{Janka93,Bauswein:2010dn,Baiotti08,Hotokezaka2011,Kiuchi2014,
DePietri2016,Endrizzi2016,Hanauske2016,Ciolfi2017,Shibata:2017b,
Radice2018,Radice2018a,Alford2018,Endrizzi2018,Kiuchi2019,DePietri2020},
in which pressure and the specific internal energy can be expressed as
the sum of a ``cold'' contribution,
obeying a zero-temperature EOS,
and of a ``thermal'' contribution obeying the ideal-fluid EOS
(see Ref.~\cite{Rezzolla_book:2013} for further details).
The latter approach, however, is still far from being realistic,
as a constant thermal adiabatic index $\Gamma_\text{th}$
does not accurately describe the behavior of nuclear matter
at finite temperature \cite{Lim2019b,Figura2020}.

We have introduced in Ref.~\cite{Lu2019} four state-of-the-art
finite-temperature EOSs constructed in the Brueckner-Hartree-Fock (BHF) approach,
which have been shown to fulfill all current constraints
imposed by observational data from nuclear structure, heavy-ion collisions,
NS global properties, and recently NS merger events \cite{Li2008b,Wei2020}.
One of those EOSs, labeled V18
(see Sec.~II for details),
was already examined in our previous paper \cite{Figura2020},
in which the merger simulations were performed mainly
employing the widely used hybrid-EOS approach.
Here we discuss novel results recently obtained from merger simulations
in which these four BHF finite-temperature EOSs were employed,
thus overcoming the approximate hybrid-EOS approach.
We find that differences observed in the simulations
are strongly related to the stiffness of the adopted EOSs.
We investigate both
the hydrodynamic and GW properties,
focusing on the stability and mass distribution of the remnant
and the properties of the ejected matter.
Since present GW detectors are not capable to see the postmerger phase of BNSMs,
our results will serve as predictions to confront with future observations.

The article is organized as follows.
We first review in Sec.~\ref{s:eos} the computation of our EOSs
in the BHF formalism and discuss their basic characteristics.
The specific properties of the GW signal to be analyzed are introduced
in Sec.~\ref{s:gws}.
The numerical setup and methods used in this work for BNSM simulations
are introduced in Sec.~\ref{s:sim}.
Results of the simulations are presented in Sec.~\ref{s:res},
and conclusions are drawn in Sec.~\ref{s:end}.

\section{Equation of state at finite temperature}
\label{s:eos}

\subsection{The microscopic BHF approach}

The extension of the BHF approach to finite temperature
was first formulated by Bloch \& De Dominicis \cite{Bloch1958}.
In the following we only provide a brief overview of the formalism
for asymmetric nuclear matter,
referring to the relevant references
\cite{Bloch1958,Lejeune1986,Baldo1999,Baldo1999a,
Nicotra2006a,Nicotra2006b,Li2010,Burgio2011,Burgio2010},
in particular the recent \cite{Lu2019},
for further details.
In this approach,
the essential ingredient is the two-body in-medium scattering matrix $K$,
which, along with the single-particle (s.p.) potential $U$,
satisfies the self-consistent equations
\be
  K(n_B,x_p;W) = V + V \;\text{Re} \sum_{1,2}
 \frac{|12 \rangle (1-n_1)(1-n_2) \langle 1 2|}
 {W - e_1-e_2 +i0} K(n_B,x_p;W) \:
\ee
and
\be
 U_1(n_B,x_p) = {\rm Re} \sum_2 n_2
 \langle 1 2| K(n_B,x_p;e_1+e_2) | 1 2 \rangle_a \:,
\ee
where $n(k)$ is a Fermi distribution,
$x_p=n_p/n_B$ is the proton fraction, and
$n_p$ and $n_B$ are the proton and the total baryon number densities,
respectively.
(In the following we also use the notation
$\rho_i=m_N n_i$ and $\rho=m_N n_B$
for the rest-mass densities,
being $m_N = 1.67\times10^{-24}\,$g the nucleon mass).
$W$ is the starting energy and
$e(k) \equiv k^2\!/2m + U(k)$ is the s.p.~energy.
The multi-indices 1,2 denote in general momentum, isospin, and spin.

Several choices for the realistic nucleon-nucleon interaction $V$
are adopted in the present calculations:
the Argonne $V_{18}$ \cite{Wiringa95},
the Bonn B (BOB) \cite{Machleidt1987,Machleidt1989},
and the Nijmegen 93 (N93) \cite{Nagels1978,Stoks1994},
and compatible three-body forces (TBF) as input.
We remind the reader that in our approach
the TBF are reduced to an effective two-body force
and added to the bare potential $V$,
see Refs.~\cite{Grange1989,Zuo2002,Li2008a,Li2008b} for details.
More precisely, the BOB and N93 are supplemented with microscopic TBF
employing the same meson-exchange parameters as the two-body potentials
\cite{Grange1989,Zuo2002,Li2008a,Li2008b},
whereas $V_{18}$ is combined either with a microscopic or a phenomenological TBF,
the latter consisting of an attractive term due to two-pion exchange
with excitation of an intermediate $\Delta$ resonance,
and a repulsive phenomenological central term
\cite{Carlson1983,Schiavilla1986,Baldo1997,Zhou2004}.
They are labeled as V18 and UIX, respectively,
throughout the paper and in all figures.

\begin{figure*}[t]
\vspace{-6mm}
\centerline{\includegraphics[scale=0.6]{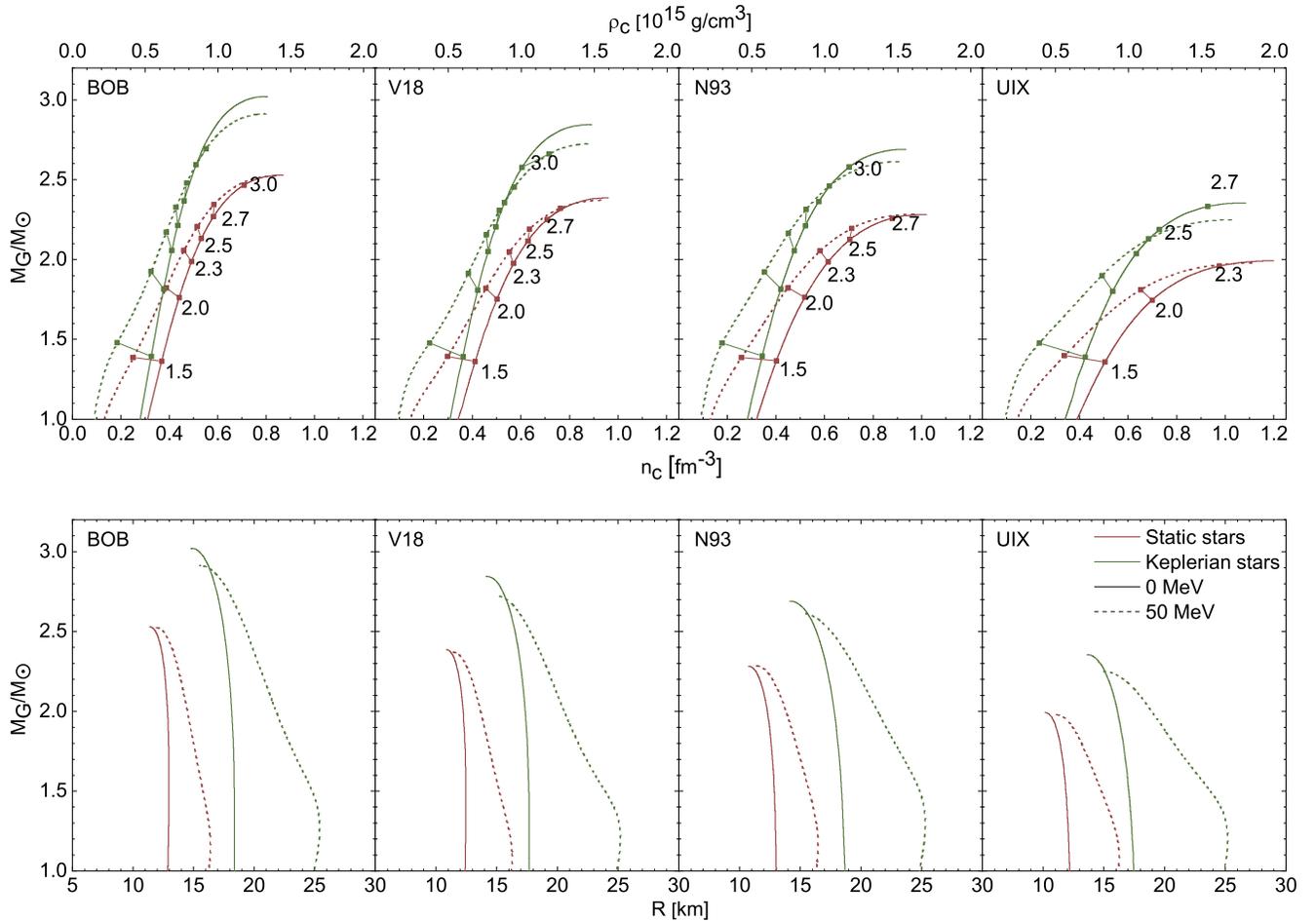}}
\vspace{-9mm}
\caption{
Gravitational mass vs central density (top row) and radius (bottom row)
for static (red curves)
and uniformly rotating at $f_\text{Kepler}$ (green curves) stars
at $T=0$ (solid curves) and $T=50\mev$ (dashed curves).
Configurations of constant baryonic masses $M_B/\ms$ (numbers)
are indicated by markers and connected by thin lines.
}
\label{f:xt}
\end{figure*}

A simplification of the Bloch \& De Dominicis scheme
can be achieved by disregarding the effects of finite temperature
on the s.p.~potential $U_{n,p}(k)$,
and using the $T=0$ results in order to speed up the calculations.
This is the so-called frozen-correlations approximation,
and it has been frequently used in some of our previous papers
\cite{Baldo1999a,Nicotra2006a,Nicotra2006b,Li2010,Burgio2011,Lu2019}.
It has been shown that at not too high temperature
($T\lesssim30\mev$)
this produces a negligible effect
on thermodynamic properties of nuclear matter
\cite{Lejeune1986,Baldo1999a,Baldo1999,Nicotra2006a,Burgio2010}.
Within this approximation,
the nucleonic free energy density has the following simplified expression,
\be
 f_N = \sum_{i=n,p} \left[ 2\sum_k n_i(k)
 \left( {k^2\over 2m_i} + {1\over 2}U_i(k) \right) - Ts_i \right] \:,
\ee
where $i$ denotes the nucleonic species and
\be
 s_i = - 2\sum_k \Big( n_i(k) \ln n_i(k) + [1-n_i(k)] \ln [1-n_i(k)] \Big)
\ee
is the entropy density
treated as a free Fermi gas with spectrum $e_i(k)$.
From the total free energy density $f=f_N+f_L$,
including the lepton contribution $f_L$ as a finite-temperature Fermi gas,
one can compute all relevant observables in a thermodynamically consistent way.
In fact one can define the chemical potentials
\be
 \mu_i = \frac{\partial f}{\partial n_i} \:,
\ee
from which the composition of betastable stellar matter can be obtained,
and then the total pressure $p$ and the internal energy density $\eps$,
\bal
 p &= n_B^2 {\partial{(f/n_B)}\over \partial{n_B}}
 = \sum_i \mu_i n_i - f \:,
\label{e:eosp}
\\
 \eps &= f + Ts \:,\quad
 s = -{{\partial f}\over{\partial T}} \:.
\label{e:eose}
\eal

In order to simplify the calculations employing these EOSs,
in Ref.~\cite{Lu2019} we gave numerical parametrizations
for the free energy density
of symmetric nuclear matter (SNM) and pure neutron matter (PNM),
and used a parabolic approximation for the $x_p$ dependence
of asymmetric nuclear matter
\cite{Burgio2010,Zuo2004,Bombaci1991,Zuo1999},
\bal
 f(n_B,T,x_p) \approx& f_\text{SNM}(n_B,T)
\\&\nonumber
 + (1-2x_p)^2 \big[ f_\text{PNM}(n_B,T) - f_\text{SNM}(n_B,T) \big] \:.
\eal
In Ref.~\cite{Fan21} it has been confirmed
that this is an excellent approximation for our purposes.
This specifies the EOS for arbitrary values of
baryon density, proton fraction, and temperature,
which can then be employed
for computing the mass-radius relation of cold NSs by
solving the Tolman-Oppenheimer-Volkov (TOV) equations
for charge-neutral betastable matter including leptons,
and in the merger simulations discussed in the following sections.

We notice that,
since our EOSs account only for homogeneous matter in the core region of the NS,
we have to attach an EOS for the crust which takes properly into account
clusterized matter at lower density
$\rho \lesssim 10^{14}\gc3$
for every value of temperature and proton fraction;
for that purpose we choose the Shen EOS \cite{Shen11}.
Furthermore, we also include in all our simulations an artificial
low-density background atmosphere,
$\rho \lesssim 10^3\gc3$,
evolved as discussed in \cite{Radice2013c}.

\begin{table*}[t]
\caption{
Properties of the maximum mass configurations of static
(first row for each EOS)
and maximally rotating stars with Kepler frequency (second row)
at temperatures $T=0$ and $50\mev$ (values in brackets):
gravitational and baryonic masses $M$ and $M_B$,
radius $R$, and central density $n_c$.
Also listed for comparison and later use are
the estimated threshold gravitational mass $M_\text{th}$
according to Eqs.~(\ref{e:mth},\ref{e:mthbaus0},\ref{e:mthbaus}),
the baryonic mass of the merger simulations, Eq.~(\ref{e:mgw2}),
the baryonic mass of the GW170817 object, Eq.~(\ref{e:mgw}),
the tidal deformability of the individual NSs,
and the radius of a cold static NS with mass $1.6\ms$.
}
\def\mr#1{\multirow{2}{*}{#1}}
\def\myc#1{\multicolumn{1}{c|}{$#1$}}
\setlength{\tabcolsep}{2.4pt}
\renewcommand{\arraystretch}{1.2}
\begin{ruledtabular}
\begin{tabular}{lcccccccccc}
  EOS & $f$ [kHz] &
  $M/\ms$ & $M_\text{th}/\ms$ & $M_B/\ms$ & $\msim/\ms$ & $\mgw/\ms$ &
  $R\;$ [km] & $n_c$ [$\fm3$] & $\la_{1.35}$ & $R_{1.6}\;$ [km] \\
\hline\mr{BOB}
 &0          &2.53 (2.52)&\mr{3.17, 3.33, 3.39}&3.10 (2.99)&\mr{2.95}&\mr{3.00}&11.38 (11.93)&0.87 (0.84)&\mr{755}&\mr{12.98}\\
 &1.70 (1.55)&3.02 (2.91)&         &3.61 (3.33)&         &         &14.86 (15.56)&0.80 (0.79)&&\\
\hline\mr{V18}
 &0          &2.39 (2.37)&\mr{3.03, 3.15, 3.23}&2.91 (2.79)&\mr{2.97}&\mr{3.01}&10.86 (11.40)&0.96 (0.93)&\mr{597}&\mr{12.45}\\
 &1.77 (1.61)&2.85 (2.73)&         &3.38 (3.10)&         &         &14.20 (14.86)&0.88 (0.89)&&\\
\hline\mr{N93}
 &0          &2.28 (2.28)&\mr{2.99, 3.14, 3.23}&2.73 (2.64)&\mr{2.94}&\mr{3.00}&10.72 (11.38)&1.02 (0.97)&\mr{701}&\mr{12.80}\\
 &1.73 (1.54)&2.69 (2.61)&         &3.15 (2.94)&         &         &14.15 (15.15)&0.93 (0.90)&&\\
\hline\mr{UIX}
 &0          &1.99 (1.98)&\mr{2.80, 2.80, 2.89}&2.35 (2.24)&\mr{2.95}&\mr{3.02}&10.16 (11.08)&1.20 (1.11)&\mr{434}&\mr{11.76}\\
 &1.72 (1.46)&2.36 (2.25)&         &2.73 (2.49)&         &         &13.61 (15.01)&1.08 (1.03)&&\\
\end{tabular}
\end{ruledtabular}
\label{t:eos}
\end{table*}

\subsection{EOS and stellar structure}
\label{s:gt}

To illustrate the difference between the four EOSs
regarding the bulk properties of NSs,
Fig.~\ref{f:xt} shows the NS gravitational mass
vs.~central density and NS radius diagrams,
obtained in the standard way by solving the TOV equations for betastable and
charge-neutral matter,
at the two temperatures $T=0,50\mev$
for both static and fastest uniformly rotating
(with mass-shedding frequency $f_\text{Kepler}$) configurations
for the different EOSs.
In the figure we also indicate by markers the baryonic masses
for different configurations.
The values of maximum masses and Kepler frequencies are also summarized
in Table~\ref{t:eos}.

Regarding the properties of the static cold NSs,
the maximum masses of all EOSs except the UIX
are larger than the current observational lower limit
$M>2.14^{+0.10}_{-0.09}\ms$ \cite{Cromartie2019}.
Concerning the radius, we found in \cite{Burgio2018,Wei2019} that
the values of a 1.4-solar-mass NS,
$R_{1.4}=12.97,12.47,12.91,11.96\,$km
for BOB,V18,N93,UIX,
fulfill the constraint derived from
the tidal deformability in the GW170817 merger event,
$R_{1.36}=11.9\pm1.4\;$km \cite{Abbott2018b}.
They are also compatible with estimates of the mass and radius
of the isolated pulsar PSR J0030+0451 recently
observed by NICER,
$M=1.44^{+0.15}_{-0.14}\ms$ and $R=13.02^{+1.24}_{-1.06}\,$km
\cite{Miller_2019}, or
$M=1.36^{+0.15}_{-0.16}\ms$ and $R=12.71^{+1.14}_{-1.19}\,$km
\cite{Riley2019}.

As seen in Fig.~\ref{f:xt} and reported in \cite{Lu2019},
the dependence of the maximum gravitational mass of static NSs on temperature
is very weak,
due to a strong compensation between nucleonic and leptonic contributions
to the thermal pressure of betastable matter in the BHF approach.
However, finite temperature decreases notably the stability
(Kepler frequencies and maximum masses)
of fast-rotating stars for all EOSs.
This is important for the analysis of BNSMs,
as essential features of the merger remnant are high temperature
($> 50\mev$)
and very fast rotation ($>1\;$kHz),
as will be illustrated later.
In fact, a merger remnant is expected to be rotating differentially
with even higher frequencies than $f_\text{Kepler}$,
and this allows a metastable transient state before collapse to a black hole
with still higher threshold mass $M_\text{th}$ than the one of rigid rotation,
according to the approximate universal relations found in
\cite{Koeppel2019,Bauswein2020a},
where the threshold mass is related to the
maximum mass of the static model $\mtov$
and one other static NS parameter,
\bal
 M_\text{th} &=
 \Big( 3.06 - \frac{1.01}{1-1.34\mtov/R_\text{TOV}} \Big) \mtov \:,
\label{e:mth}
\\
 M_\text{th} &= 0.59 \mtov + 1.36\ms + 0.80\ms\,\Lambda_{1.4}/1000 \:, 
 \label{e:mthbaus0}
\\
 M_\text{th} &= 0.55 \mtov - 0.20\ms + 0.17\ms R_{1.6}/\text{km} \:. 
\label{e:mthbaus}
\eal
These estimates are also listed in Table~\ref{t:eos}
and might be up to about 20\% larger than the ones for rigid rotation
for the softest UIX EOS.
However, the predictions themselves vary by several percent,
the latter values, Eq.~(\ref{e:mthbaus}),
being significantly higher for most EOSs.

Thus finite temperature and rotation have opposite effects
on the stellar stability
and their competition determines the stability limit of a BNSM event,
for example.
That is why an accurate theoretical determination of the finite-temperature EOS
is essential for the analysis of a merger event.
No firm conclusions regarding properties of cold NSs can be drawn
based on an analysis of a hot merger remnant
unless this feature is well under theoretical control.

For the specific case of the GW170817 event,
an important quantity is its total baryonic mass
\be
 \mgw \equiv 2 M_B(M_G=1.365\ms) \:,
\label{e:mgw}
\ee
which we also list in the table
together with the relevant value for the simulations we actually carried out,
\be
 \msim \equiv 2 M_B(M_G=1.35\ms) \:.
\label{e:mgw2}
\ee
It depends only very weakly on the EOS.
Comparing the values of $M_B$, $\msim$, and $\mgw$,
we can already draw some important qualitative conclusions:
While the BOB, V18, and (marginally) N93 EOSs would be able to sustain even
a rigidly rotating hot remnant,
the soft UIX EOS would permit only a metastable differentially rotating one.
Eventually, the cooling-down remnant would gain stability
(not enough for UIX though),
but in the long-term spindown,
only the BOB EOS would be able to sustain a stable cold and static NS
with a mass of $\msim$ or $\mgw$
(All this assuming that no mass is ejected).
These are very simplistic considerations
that we will confront now with our results of the merger simulations.

\section{Gravitational-wave signal}
\label{s:gws}

As a standard approach in numerical relativity,
we adopt the Newman-Penrose formalism \cite{Newman62a}
in order to extract the GW strains for our models.
In particular, the Einstein toolkit module \textsc{WeylScal4}
is used in order to calculate the Newman-Penrose scalar $\psi_4$
at different surfaces of constant coordinate radius $r$.
$\psi_4$ is then related to the second time derivatives
of the GW polarization amplitudes $h_+$ and $h_\times$ via
\be
 \psi_4 = \ddot{h}_+ - i\ddot{h}_\times =
 \sum_{l=2}^\infty \sum_{m=-l}^l
 \psi_4^{\ell m}(t,r)\; _{-2}Y_{\ell m}(\theta,\phi) \:,
\ee
where we adopt the double-dot notation in order to express
the second time derivative
and we have also considered the multipole decomposition of $\psi_4$
in spherical harmonics \cite{Goldberg:1967} of spin weight $s=-2$;
in our numerical setup,
such decomposition is carried out by the module \textsc{Multipole}.
We restrict our analysis to the $\ell=m=2$ mode,
which represents the dominant one after the merger;
in particular, we assume
\be
 h_{+,\times} = \sum_{l=2}^\infty \sum_{m=-l}^l
 h_{+,\times}^{\ell m}(t,r)\; _{-2}Y_{\ell m}(\theta,\phi)
 \approx h^{22}_{+,\times}(t,r)\; _{-2}Y_{22}(\theta,\phi) \:.
\label{e:hpol}
\ee
The double integration in time
of $\psi_4$ is performed according to the
fixed-frequency integration method described in \cite{Reisswig:2011}.
Our waveforms are then aligned to the ``time of the merger"
(as done, e.g., in Ref.~\cite{Rezzolla2016}),
which we impose as $t=0$ and define as the time when the GW amplitude
\be
 |h| \equiv \sqrt{h^2_+ + h^2_\times}
\label{e:hmod}
\ee
reaches its global maximum.
In our analysis,
we also compute the instantaneous frequency of the GWs,
defined as in \cite{Read2013},
\be
 f_\text{GW} \equiv \frac{1}{2\pi}\dv{\chi}{t} \:,
\ee
where $\chi$ = arctan$(h_\times / h_+)$
represents the phase of the complex waveform.
As done in Ref.~\cite{Rezzolla2016},
we identify
\be
 f_\text{max} \equiv f_\text{GW}(t=0)
\label{e:fmax}
\ee
as the instantaneous frequency at amplitude maximum.

An important quantity in our analysis is the
power spectral density (PSD) of the effective amplitude,
\be
 \tilde{h}(f) \equiv
 \sqrt{\frac{\abs{\tilde{h}_+(f)}^2 + \abs{\tilde{h}_\times(f)}^2}{2}} \:,
\label{e:psdeq}
\ee
where $\tilde{h}_{+,\times}(f)$ represent the Fourier transforms of
$h_{+,\times}$, respectively,
\be
 \tilde{h}_{+,\times}(f) \equiv
 \int dt e^{-i2\pi ft} h_{+,\times}(t)
\label{e:hpxt}
\ee
for $f \geq 0$,
and $\tilde{h}_{+,\times}(f) \equiv 0$ for $f<0$.
Our PSDs are first filtered by applying a symmetric time-domain Tukey filter
with parameter $\alpha=0.25$ to the waveforms,
in order to compute PSDs without the artificial noise due to
the truncation of the waveforms themselves.
We then focus on determining the $f_2$ peak of the PSD;
in this regard, we first fit our data with the
analytic function \cite{Takami2015}
\be
 S_2(f) = A_{2G} e^{-(f-F_{2G})^2/W_{2G}^2} + A(f) \gamma(f) \:,
\ee
where
\bal
 A(f) &\equiv \frac{1}{2W_2}
 \qty[ (A_{2b}-A_{2a})(f-F_2) + W_2(A_{2b}+A_{2a}) ] \:,
\\
 \gamma(f) &\equiv
 \qty( 1+e^{-(f-F_2+W_2)/s} )^{-1}
 \qty( 1+e^{(f-F_2-W_2)/s} )^{-1} \:.
\eal
The peak frequency is then determined by
\be
 f_2 \equiv \frac{\int df\,S_2(f)\,f}{\int df\,S_2(f)} \:.
\label{e:f2}
\ee
An intrinsic uncertainty,
due to both the choice of the fitting functions and parameters,
and the integration interval,
affects the fitting procedure, and
we estimate the latter as $\pm 10\,\rm{Hz}$;
this estimate is later added in quadrature to a systematic deviation
of the value we find for $f_2$ from the nearest (local) maximum of the PSD curve,
which in all our cases also coincides with the global maximum.

We complete our analysis with the calculation of the total emitted energy
for the $\ell=m=2$ mode, namely
\be
 E_\text{GW} = \frac{R^2}{16\pi} \int dt \int d\Omega\;
 \abs{\dot{h}(t,\theta,\phi)}^2 \:,
\label{e:egw}
\ee
where $\Omega$ labels the solid angle and $R$ represents the
source-detector distance.

\begin{figure*}[t]
\vspace{-0mm}
\centerline{\hspace{5mm}\includegraphics[scale=0.31]{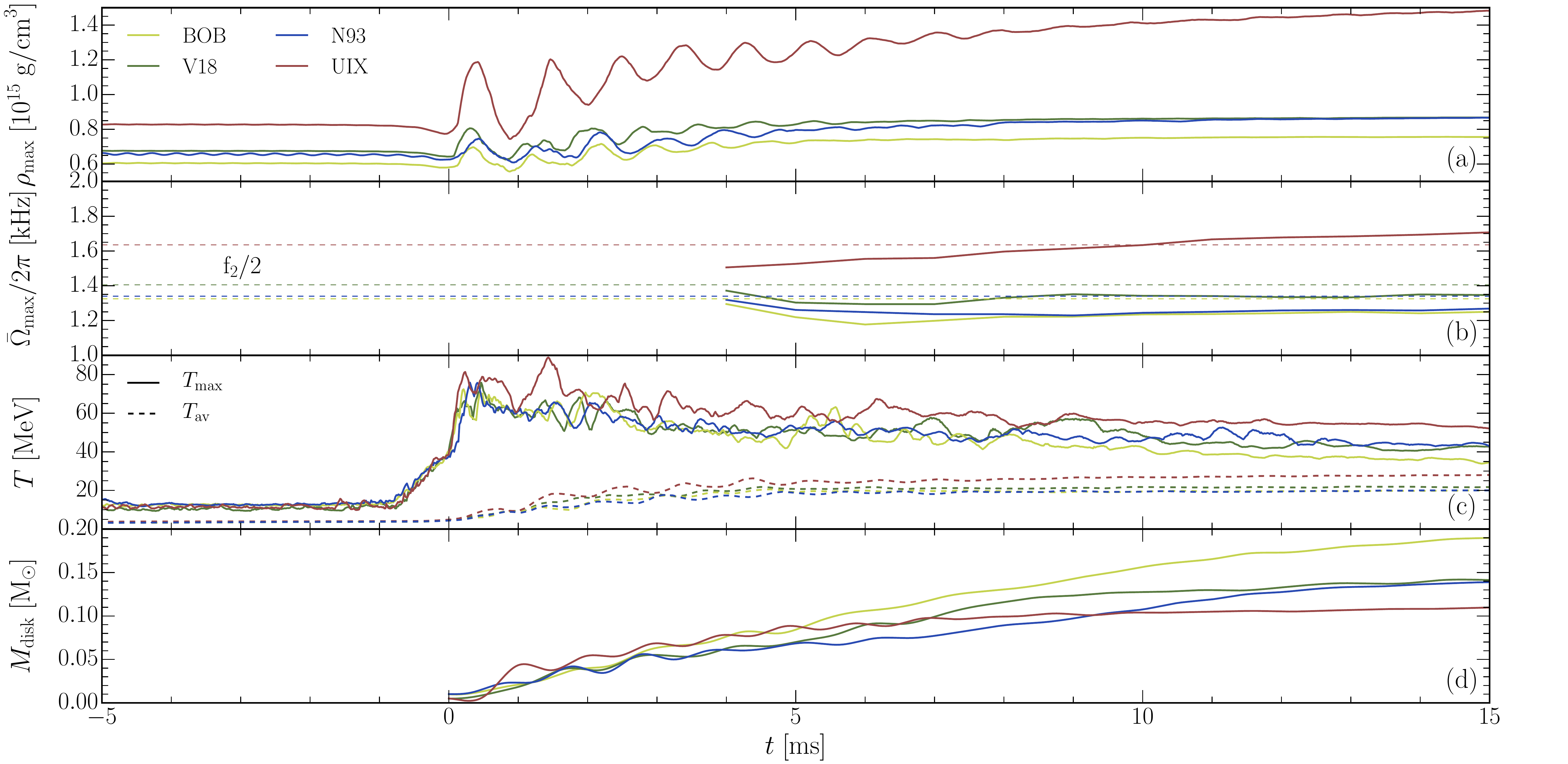}}
\vspace{-3mm}
\caption{
Maximum values of rest-mass density (a)
and azimuthally-averaged differential rotational frequency (b),
maximum $T_\text{max}$ and average $T_\text{av}$ temperature (c),
and disk mass (d)
as a function of time
for the simulations using the four different EOSs.
}
\label{f:maxrhot}
\end{figure*}

\section{Merger Simulations}
\label{s:sim}

Before illustrating the results of the merger simulations,
we briefly review the mathematical and numerical setup that we adopt.
This is similar to the one of Ref.~\cite{Figura2020}
and discussed in great detail in Ref.~\cite{Papenfort2018},
to which we refer the interested reader for additional information.

Our models employ initial data for irrotational binary NSs computed
using the multi-domain spectral-method code LORENE \cite{Lorene,Gourgoulhon01}.
In our case all initial data,
modeled considering a zero-temperature, beta-equilibrated cut
of the full EOS tables,
involve equal-masses binaries with a gravitational mass $M_G=1.35\ms$
at infinite separation
(corresponding to a total baryonic mass $\msim\approx2.94-2.97\ms$
slightly dependent on the EOS, given in Table~\ref{t:eos}),
and an initial separation between the stellar centers of $45\;$km.
We stress that with this choice our simulations can reproduce closely the
GW170817 merger event,
in which the detected chirp mass $M_\text{chirp}=1.188\ms$
corresponds to $M_G=1.365\ms$
for a symmetric binary system \cite{Abbott2017_etal}.

We perform our simulations in full general relativity
using for the spacetime evolution
the fourth-order finite-differencing McLachlan code \cite{Brown:2008sb},
part of the publicly available Einstein toolkit \cite{loeffler_2011_et},
with the inclusion of a fifth-order Kreiss-Oliger-type artificial dissipation
\cite{Kreiss73} to ensure the non-linear stability of the evolution.
In particular,
our simulations adopt the CCZ4 formulation of the Einstein equations
\cite{Alic:2011a,Alic2013,Bezares2017},
where the evolution of the gauge variables is carried out by using a
``1+log'' slicing condition for the lapse function and a
``Gamma driver'' condition for the shift vector
(see, e.g., Refs.~\cite{Alcubierre2003, Pollney:2007ss:shortal}).
In our setup,
the WhiskyTHC code \cite{Radice2013b,Radice2013c,Radice2015} is used
in order to solve the general-relativistic hydrodynamics equations;
in particular, the latter employs either finite-volume or high-order
finite-differencing high-resolution shock-capturing methods and,
for our simulations,
we adopt the HLLE Riemann solver and the high-order MP5 primitive reconstruction
\cite{suresh_1997_amp,Radice2012a}.
The coupled set of the spacetime and hydrodynamic equations is integrated in time
using the method of lines with an explicit third-order Runge-Kutta method,
where a Courant-Friedrichs-Lewy (CFL) parameter of 0.15 is used
in order to compute the timestep.
Regarding our grid setup,
we employ the Carpet driver \cite{Schnetter-etal-03b},
which operates, with an adaptive-mesh-refinement approach,
the following grid hierarchy:
we consider six refinement levels with a grid resolution which ranges from
$\Delta h_5 = 0.16\ms$ (i.e., $\sim 236\,$m) for the finest level to
$\Delta h_0 = 5.12\ms$ (i.e., $\sim 7.5\,$km) for the coarsest level,
whose outer boundary is placed at $1024\ms$ (i.e., $\sim 1515\,$km).
Our setup also makes use of a reflection symmetry across the $z=0$ plane
in order to reduce the computational resources needed.

Neutrino emission acts as cooling mechanism and is implemented
in our temperature-dependent simulations.
We treat the effects on matter due to weak reactions
using the gray (energy-averaged) neutrino-leakage scheme
described in Refs.~\cite{Galeazzi2013,Radice2016},
and evolve free-streaming neutrinos according to the M0 heating scheme
introduced in Refs.~\cite{Radice2016,Radice2018a}.
This is accompanied by a loss of betastability of the heated stellar matter
that was analyzed in detail in Ref.~\cite{Figura2020}.

Our simulations do not include a treatment for viscous effects,
although it has been shown (see
Refs.~\cite{Shibata:2017b,Kiuchi2017,Fujibayashi2017,Radice2018a,DePietri2020}
for a complete discussion)
that the latter have an impact on several features of the remnant,
such as the angular velocity distribution,
the emitted GW signal,
and ejecta properties.
As a consequence, and as better investigated in the next chapter,
we do not expect a significant slowdown of rotation
in the timespan we consider in our simulations.

Our EOS tables cover a range
$5.1 \leq \log_{10}(\rho/\gc3) \leq {16}$
in rest-mass densities,
with a spacing $\Delta\log_{10}(\rho/\gc3) = 0.1$,
for a total of 110 points;
temperature ranges from
$-1.0 \leq $ log$_{10}(T/\text{K}) \leq {2.6}$,
with a spacing $\Delta\log_{10}(T/\text{K})=0.04$ for a total of 91 points,
and electron fractions cover the range
$0.01 \leq Y_e \leq 0.65$,
where the spacing is $\Delta Y_e = 0.01$, for a total of 65 points.
Our tables are first prepared in the same format as the one discussed
in Appendix~A of Ref.~\cite{Shen11};
we then use the routines present in \cite{stellarcollapse}
in order to create versions of the EOSs compatible with WhiskyTHC.
The latter code contains routines in order to carry out
either linear or cubic spline interpolations on the original tables;
the code is also responsible for the time evolution
of proton and neutron number densities,
guaranteeing the local conservation of both species
(see Ref.~\cite{Radice2016} for a detailed description).

We stress that,
while the V18 EOS has already been studied in Ref.~\cite{Figura2020},
the BOB, N93, and UIX EOSs are employed here for the first time
in merger simulations.
In the following we present the results.

\section{Results and discussion}
\label{s:res}

All the simulations presented here follow the remnant evolution
for a period of at least 15 ms.
We set our time coordinate such that $t = t_\text{merg} = 0$,
where $t_\text{merg}$ is the time of the merger
and corresponds to the maximum of the GW amplitude.
We notice that for all EOSs the merger simulations produce
a metastable hypermassive NS during this time,
when the remnant is still stabilized by differential rotation
and finite temperature.
This feature is compatible with the multimessenger analysis
of the GW170817 event \cite{Gill2019}.

\subsection{Stellar matter}

\begin{figure*}[t]
\vspace{-0mm}
\centerline{\hspace{7mm}\includegraphics[scale=0.31]{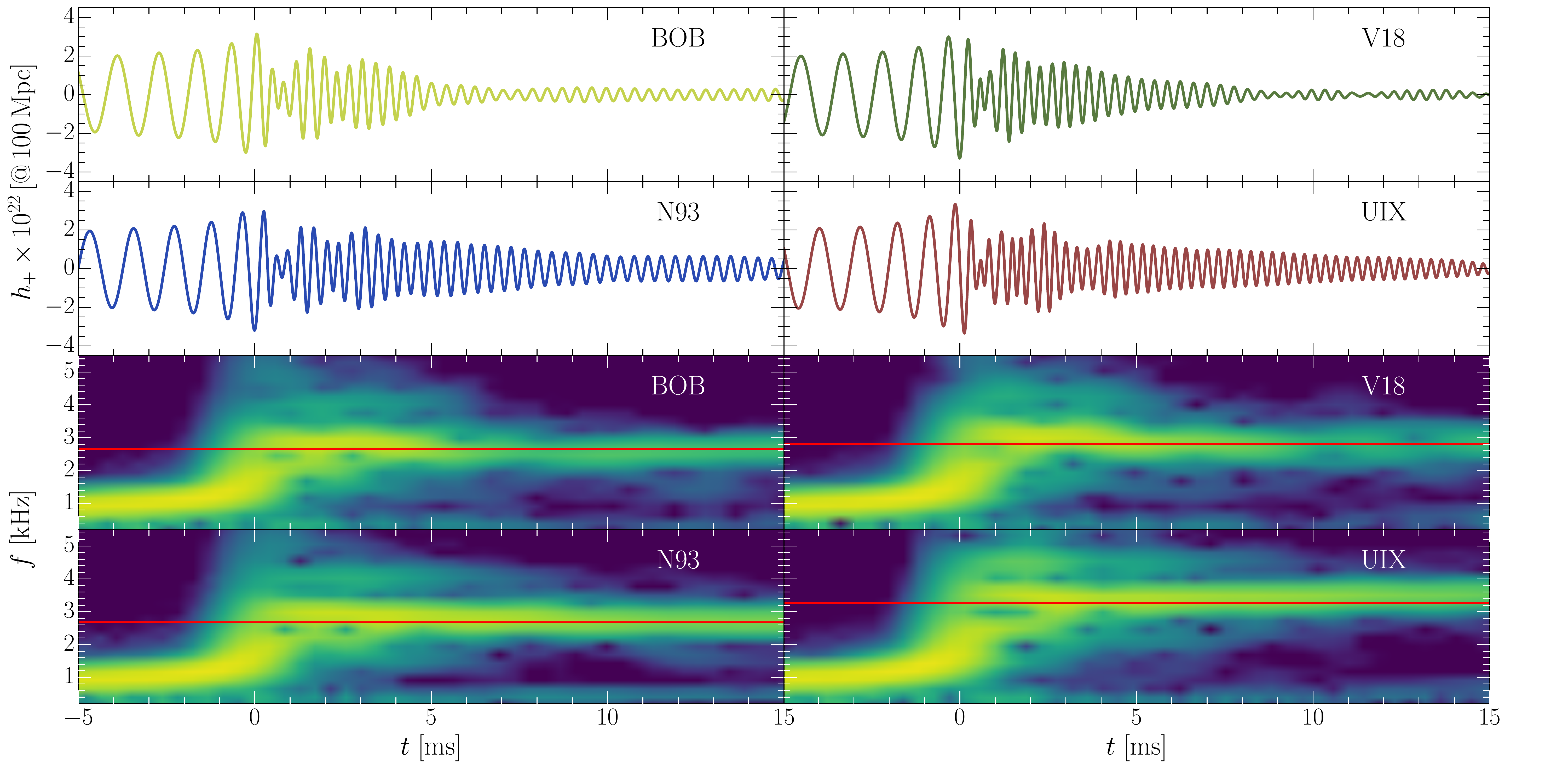}}
\vspace{-3mm}
\caption{
Upper panels:
gravitational waveforms over a time scale of 15 ms after the merger
for the four simulated models obtained for gravitational masses $2\times1.35\ms$.
Lower panels:
the spectrograms for all the considered cases;
red lines represent the position of the $f_2$ peaks
(Table~\ref{t:sim}).
}
\label{f:strain}
\end{figure*}

As a first illustration
of the typical properties of the stellar matter in the postmerger phase,
Fig.~\ref{f:maxrhot} shows for the different cases we have studied
the evolution of the maximum rest-mass density 
$\rho_\text{max}$ (a),
the maximum azimuthally-averaged differential rotation frequency
$\bar\Omega_\text{max}/2\pi$ (b),
the maximum and averaged temperatures $T_\text{max}$ and $T_\text{av}$ (c),
the latter quantities evaluated in the $z=0$ plane,
and the mass of the disk (d).
We now discuss the results in detail.

We find that the simulations performed with the most realistic V18 and N93 EOSs
lead to a remnant with $\rho_\text{max}$ of about $0.9\times10^{15}\gc3$,
and also similar values of the maximum and average temperatures.
These two EOSs feature also similar common properties
for the static and Kepler configurations, see Table~\ref{t:eos}.
Consistently,
the post-merger remnant modeled with the BOB EOS,
which is the stiffest EOS in our sample,
reaches the smallest maximum density and temperature,
whereas the (too) soft UIX case exhibits the typical increasing central density
signature of a model experiencing a collapse after the merger
(although not within our simulation timespan),
which would be in agreement with the characteristics of this EOS
discussed in Sec.~\ref{s:gt}.
Namely, we notice that the maximum mass of the UIX Keplerian configurations,
reported in Table~\ref{t:eos},
is well below the mass of the remnant,
and therefore it might be only temporarily supported by differential rotation.
The subsequent collapse would require first a slowdown of the stabilizing
differential rotation,
which should occur on a typical time scale of milliseconds
for a too soft EOS like UIX
\cite{Hanauske2016,Radice2017b,Figura2020,Bernuzzi2020},
although in our simulation we did not detect it within 20 ms.

The determination of the actual, much longer,
collapse time of the GW170817 remnant is a very delicate task,
since it depends on several physical processes, e.g.,
the time evolution of the differential rotation \cite{Kastaun2016,Hanauske2016},
ejection of matter \cite{Rosswog1999,Radice2016,Bovard2017}, and
viscosity effects \cite{Shibata2017b,Radice2017,Alford2018}.
A tentative approach has been recently discussed in Ref.~\cite{Gill2019},
where the properties of the kilonova emission \cite{Abbott2017b}
have been combined with the delay time between the GW chirp signal
and the prompt gamma-ray emission onset in GRB 170817A \cite{Abbott2017d},
in order to estimate a collapse time of the HMNS of about 1 second.
Realistic EOSs like V18 or N93,
combined with an elaborate simulation procedure taking into account
with sufficient accuracy all the above (micro)physics ingredients,
would be expected to predict compatible values.
At the current stage we are still far from this situation.

\begin{figure*}[t]
\vspace{-2mm}\hspace{6mm}
\includegraphics[scale=0.28]{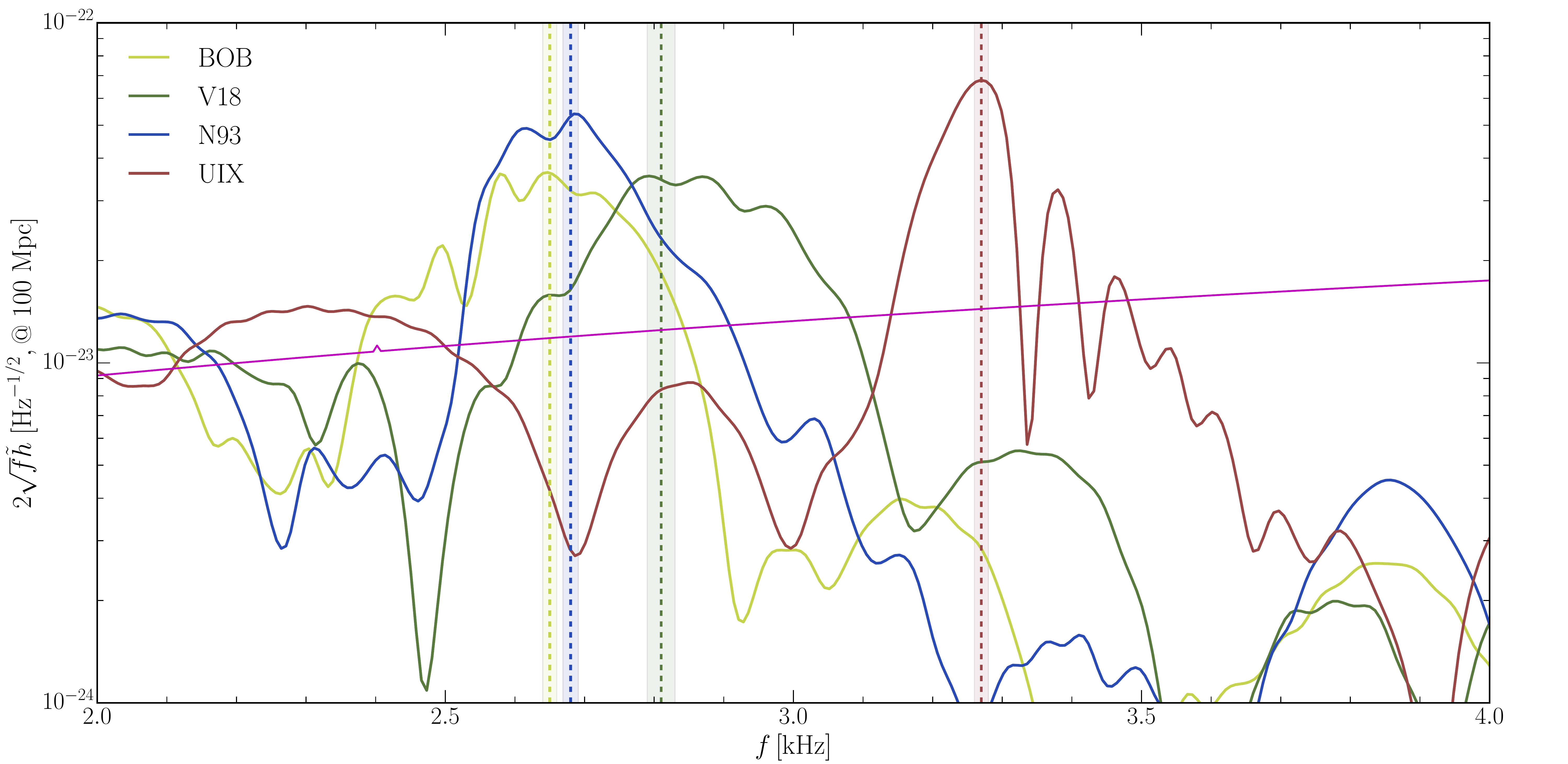}
\vspace{-3mm}
\caption{
PSDs $\tilde{h}$, Eq.~(\ref{e:psdeq}),
of the simulations evaluated at a distance of 100 Mpc.
Vertical dashed lines of different colors indicate the frequency
of the main postmerger peak $f_2$ and estimated error.
The sensitivity curve (magenta color) of Advanced LIGO
is displayed for reference.
}
\label{f:psd}
\end{figure*}

\begin{figure}[t]
\vspace{-1mm}
\centerline{\hspace{2mm}\includegraphics[scale=0.45]{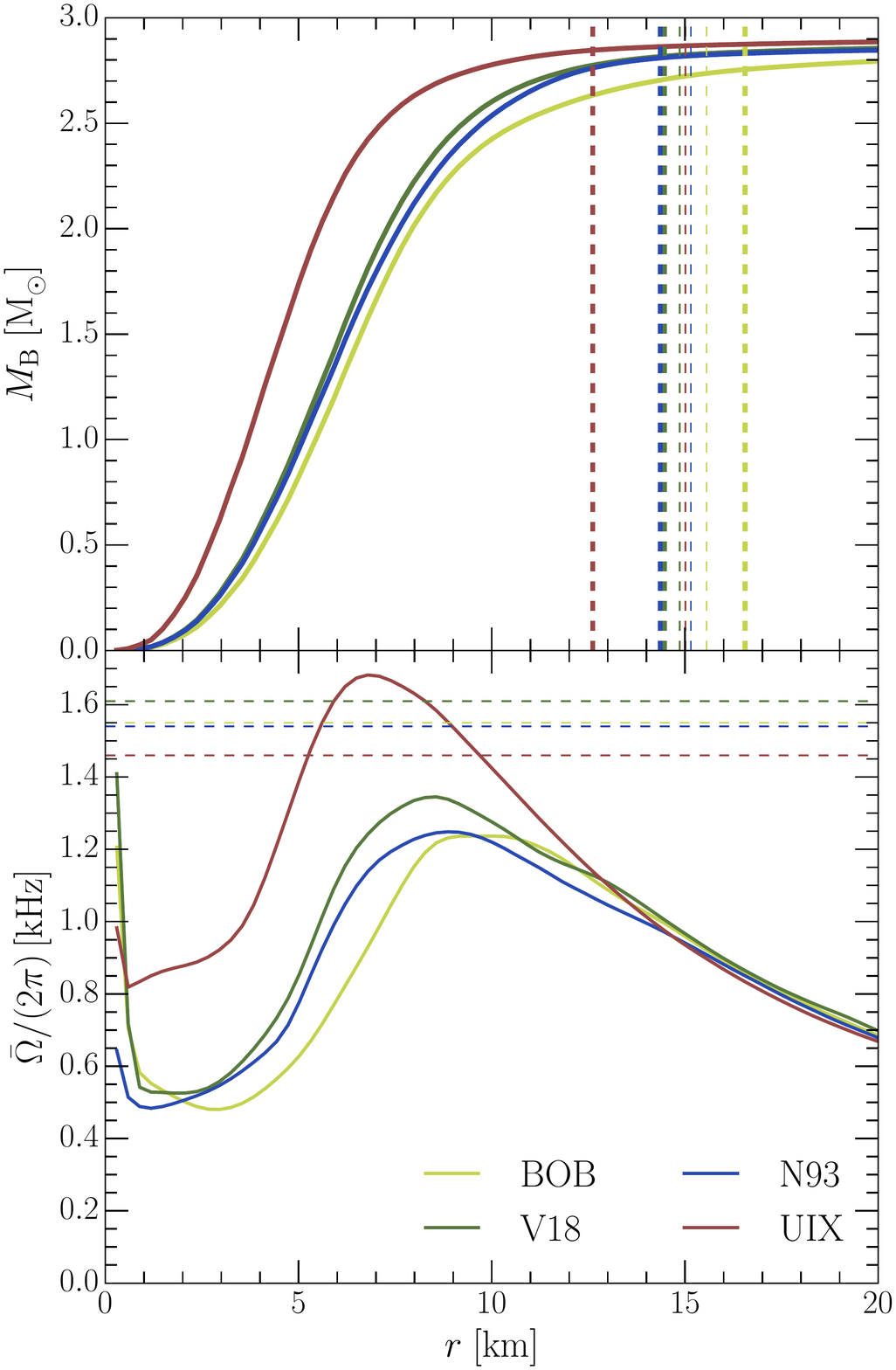}} 
\vspace{-3mm}
\caption{
Upper panel:
Enclosed baryonic mass $M_B$ as a function of spherical radius $r$
at $t=15\,$ms for the different EOSs.
Thick dashed lines denote radii corresponding to $M_\text{obj}$
in Table~\ref{t:sim}.
Thin dashed lines are the radii of $T=50\mev$ $\mmax$ Kepler configurations
in Table~\ref{t:eos}.
\\
Lower panel:
Azimuthally-averaged angular velocity, Eq.~(\ref{e:oav}),
as a function of the radial cylindrical coordinate $r$ (at $z=0$)
at $t=15\,$ms.
Thin horizontal dashed lines indicate the Kepler frequencies of the $T=50\mev$
$\mmax$ configurations for the different EOSs listed in Table~\ref{t:eos}.
}
\label{f:mb}
\end{figure}

We nevertheless continue our analysis of the numerical results
regarding rotational properties of the remnant.
Panel (b) of Fig.~\ref{f:maxrhot}
shows the time development of the maximum
of the azimuthally-averaged differential rotation frequency \cite{Hanauske2016}
on the equatorial plane (see Fig.~\ref{f:mb}),
\be
 \bar\Omega(r,t) \equiv
 \frac{1}{4\pi\Delta t}
 \int_{t-\Delta t}^{t+\Delta t} dt'
 \int_{-\pi}^{\pi} d\phi \;
 \Omega(z=0,r,\phi,t') \:
\label{e:oav}
\ee
with $\Delta t=0.5\;$ms.
Values are sampled at each ms starting from 4 ms after the merger,
as for earlier times the system is still too asymmetric.

Again the BOB, V18, and N93 EOS exhibit common features,
whereas UIX displays a different trend.
While the profile for the UIX EOS shows an increasing unstable behavior
in the time window analyzed here,
which is compatible with the increase of $\rho_\text{max}$ discussed before,
the other EOSs show stable profiles,
thus indicating no slowdown of rotation
within the milliseconds time interval simulated here.
Therefore an eventual collapse with these EOSs,
related to loss of stabilizing rotation,
could occur only much later.

For completeness (c.f., \cite{Hanauske2016}),
dashed horizontal lines shown in the same panel represent for each EOS
the quadrupole peak frequency $f_2/2$, Eq.~(\ref{e:f2}),
determined via the PSDs shown in the next section. 
We see that the maximum differential rotation frequencies
are systematically slightly lower than the $f_2$-related frequencies.
This is not surprising,
since the latter values are determined through PSDs
considering also the first 4 ms,
when the remnant is rotating slightly faster.

In panel (c) of Fig.~\ref{f:maxrhot}
we plot both the maximum temperature (solid curves)
and the density-weighted average temperature (dashed curves),
defined as
\be
 T_\text{av} 
 \equiv \frac{\int dV\rho\, T}{\int dV\rho} \:.
\label{e:tav}
\ee
All the simulations feature maximum temperatures which remain in general
lower than $70\mev$ in the post-merger phase.
As already shown in Ref.~\cite{Figura2020}, however,
maximum temperatures are reached only in local hot spots,
and are not representative of the average temperature of matter,
which is about $20$ to $30\mev$.
Typical temperatures depend slightly on the EOS,
with softer (stiffer) ones producing higher (lower) temperatures.

We also monitor the disk mass $\mdk$
(discussed in Sec.~\ref{s:mej})
as function of time,
shown in Fig.~\ref{f:maxrhot}(d).
We notice that during the short timespan of the simulations,
the disk masses are still increasing,
but tend to become stable at the end of our time evolution,
where they span a range comprised between 0.1 and $0.2\ms$,
as also reported in Table~\ref{t:sim}.
The softest UIX EOS attracts most material into the dense core
and produces the lightest disk,
contrary to the stiffest BOB model,
which is instead responsible of a large $\mdk\approx 0.2\ms$.

\begin{table*}[t]
\caption{
Properties of the simulated models:
frequency of the $f_2$ peak,
frequency at maximum amplitude $f_\text{max}$,
the total emitted GW energy until $t=15\,$ms $E_{\rm{GW}}$,
and baryonic masses of the object $M_\text{obj}$,
the disk $\mdk$,
the ejected matter $M_\text{ej}$
at $t=15\,$ms.
The $f_2$ values in brackets are obtained using the universal relations
Eqs.~(\ref{e:unir},\ref{e:unim},\ref{e:unil}).
The $\mdk$ values in brackets are obtained using the universal relation
Eq.~(25) of \cite{Radice2018a}.
}
\def\myc#1{\multicolumn{1}{c|}{$#1$}}
\setlength{\tabcolsep}{2pt}
\renewcommand{\arraystretch}{1.2}
\begin{ruledtabular}
\begin{tabular}{lcccccc}
 EOS & $f_2$ [kHz] & $f_\text{max}$ [kHz] & $E_\text{GW}$ [$10^{52}\;$erg] &
 $M_\text{obj}$ [$\ms$] & $\mdk$ [$\ms$] & $M_\text{ej}$ [$10^{-3}\ms$] \\
\hline
 BOB & 2.65$\pm$0.01 (2.62, 2.65, 2.82) & 1.68 & 4.10 & 2.76 & 0.189 (0.139) & 3.7 \\ 
 V18 & 2.81$\pm$0.02 (2.90, 2.86, 2.96) & 1.77 & 4.88 & 2.82 & 0.141 (0.093) & 4.2 \\ 
 N93 & 2.68$\pm$0.01 (2.67, 2.72, 2.87) & 1.68 & 5.99 & 2.81 & 0.138 (0.124) & 4.5 \\ 
 UIX & 3.27$\pm$0.01 (3.23, 3.19, 3.14) & 1.91 & 7.69 & 2.85 & 0.109 (0.044) & 7.4 \\ 
\end{tabular}
\end{ruledtabular}
\label{t:sim}
\end{table*}

\subsection{Gravitational-wave signal}

We now turn to the analysis of the GW signal.
In Fig.~\ref{f:strain} (upper panels) we show the plus polarization
of the $l=m=2$ component of the GW strains,
which we label as $h_+$, Eq.~(\ref{e:hpol}),
for all the considered simulations we have carried out
using different microscopic EOSs.
All models feature an instant of the merger
of about 15 ms from the start of the simulation,
identified as previously mentioned
as the time corresponding to the maximum strain amplitude.
One can roughly observe that the oscillations for the more stable EOSs
with higher $\mmax$ (BOB and V18)
are `ringing down' faster than for the others,
which is the expected behavior
\cite{Takami2015,Rezzolla2016}.


The evolution of the characteristic frequencies for the cases we have considered
is also evidenced in the lower panels of the figure,
where the spectrograms of the four models are shown.
In order to compute them,
we calculate the spectra by first segmenting the $h_+$ signals
in pieces of $\approx5\;$ms each;
a Blackman window is then applied to the segments,
which are overlapped by 90\%,
similarly to what is done in Ref.~\cite{Rezzolla2016}.
As clearly visible,
while shortly after the merger different significative frequencies are present,
like the $f_1$ or the $f_3$ peaks
(whose investigation is not reported here),
the $l=m=2$ frequency
(denoted in the plots with a red dashed line for each case)
and its related $f_2$ peak
(following the same nomenclature as in Ref.~\cite{Rezzolla2016})
is the only robust feature which is present from the time of the merger
to the end of the considered time window for all our models.

Fig.~\ref{f:psd} shows the power spectral density (PSD) plots
of all simulations,
determined as detailed in Sec.~\ref{s:gws}.
In particular, we choose to study the dominant $l=m=2$ mode,
and consider the position of the $f_2$ peak
as a tracker of the different behaviors.
Since, except for the UIX case,
it is difficult to distinguish the dominant $f_2$ peaks by eye,
the fitting procedure discussed in Sec.~\ref{s:gws}
represents the only way for an accurate determination of the $f_2$ positions,
which are shown in the figure together with the estimated errors.

We also report in Table~\ref{t:sim} these values,
together with other relevant GW properties for each simulation;
in particular, we determine for each case the maximum frequency $f_\text{max}$,
Eq.~(\ref{e:fmax}),
and the emitted GW energy $E_\text{GW}$ for the $l=m=2$ mode,
Eq.~(\ref{e:egw}),
both measured as outlined in Sec.~\ref{s:gws}.
We find that the UIX EOS predicts by far the highest frequency
for the $f_2$ peak,
more than $400\,$Hz higher than the other cases;
this represents a spectroscopical confirmation of this remnant
being the most dense
(see the maximum rest-mass density shown in Fig.~\ref{f:maxrhot})
and the fastest rotating of all the cases we considered,
since the frequency of the mode scales with the square root
of the average density
(see, e.g., Ref.~\cite{Kokkotas99b}).

For comparison we also list (in brackets) the values of $f_2$ according to
universal relations between $f_2$ and the radius $R_{1.6}$
of a $1.6\ms$ NS \cite{Bauswein2012},
the chirp mass $M_\text{chirp}$ \cite{Vretinaris2020},
and the tidal deformability parameter $\la$ \cite{Rezzolla2016}
proposed in different publications,
\bal
 f_2 [\text{kHz}] &\approx
 \left\{\begin{array}{ll}
 6.284 - 0.2823\,R_{1.6} & (f_2 < 2.8\;\text{kHz})
 \\
 8.713 - 0.4667\,R_{1.6} & (f_2 > 2.8\;\text{kHz})
 \end{array}\right. \:,
\label{e:unir}
\\
 f_2 [\text{kHz}] &\approx
 13.82 M_c - 0.576 M^2_c + 0.479 M^3_c
\nonumber\\ &\quad
 - 1.375 R_{1.6} M_c - 0.073 R_{1.6} M^2_c + 0.044 R^2_{1.6} M_c \:,
\label{e:unim}
\\
 f_2 [\text{kHz}] &\approx
 5.832 - 0.8\, \la^{1/5} \:,
\label{e:unil}
\eal
where $M_c\equiv M_\text{chirp}/\ms$ and $R_{1.6}$ is given in km.
Both $R_{1.6}$ and $\la$ are listed in Table~\ref{t:eos}.
One observes a reasonable agreement (within 3\%)
in particular for the first correlation with the radius,
whereas the last one with $\la$ is less pronounced
with about 7\% possible deviations, as in \cite{Rezzolla2016}.

Interestingly, as shown, e.g., in Refs.~\cite{Bauswein2010,Figura2020},
the position of the theoretical $f_2$ peak may change up to several tens Hz
when simulations are performed with the same zero-temperature EOS and initial data,
but employing the approximate hybrid finite-temperature EOS approach
with different values of the thermal index $\Gamma_\text{th}$.
We refer the interested reader to the latter references for complete discussions.

\subsection{Masses and ejecta}
\label{s:mej}

We also list in Table~\ref{t:sim}, for the instant $t=15$ ms,
the baryonic masses of the remnant $M_\text{obj}$ and its disk $\mdk$,
obtained by integrating the conserved rest-mass density
over the respective 3D domains
(We choose $10^{13}\gc3$ as the boundary density between object and disk,
which represents a common choice in literature,
see, e.g., Ref.~\cite{Bernuzzi2020b}),
\be
 M_B = \int d^3r\; D \ ,\quad
 D = \sqrt{\gamma}\; W \rho \:,
\ee
where $\gamma$ represents the 3-metric determinant
and $W$ is the Lorentz factor.
From Fig.~\ref{f:maxrhot} it seems that at $t=15\;$ms nearly stable
values have been reached \cite{Bernuzzi2020},
apart for the UIX EOS.

We remind that lower limits on the disk mass of GW170817
derived from its electromagnetic counterpart are about
$\mdk\gtrsim0.04\ms$
\cite{Radice2018a,Radice2018c,Kiuchi2019},
with which all our EOSs would comply.
An approximate universal relation between $\mdk$ and $\Lambda$
has been proposed in Ref.~\cite{Radice2018a} 
and we list those values also in Table~\ref{t:sim}.
However, it can be seen that the deviations are very large,
which has also been pointed out in \cite{Kiuchi2019}.

To measure the properties
(baryonic mass $M_\text{ej}$)
of the dynamical ejecta
(treated as perfect fluid; no nuclear reaction network is present in the code),
we consider multiple spherical detectors at different radii around the origin,
taking the detector at $200\ms \approx 300$ km for our measurements.
In order to determine which of the material crossing this surface
is effectively unbound,
we set a threshold according to the geodesic criterion,
as done, for example, in Refs.~\cite{Papenfort2018,Bernuzzi2020}.
In detail, a particle on geodesics is considered to be unbound
if the covariant time component of the fluid four-velocity $u$ satisfies
$u_t \le -1$
(see, e.g., Ref.~\cite{Bovard2017} for a discussion of the method).

\begin{figure*}[t]
\vspace{-1mm}
\centerline{\includegraphics[scale=0.35]{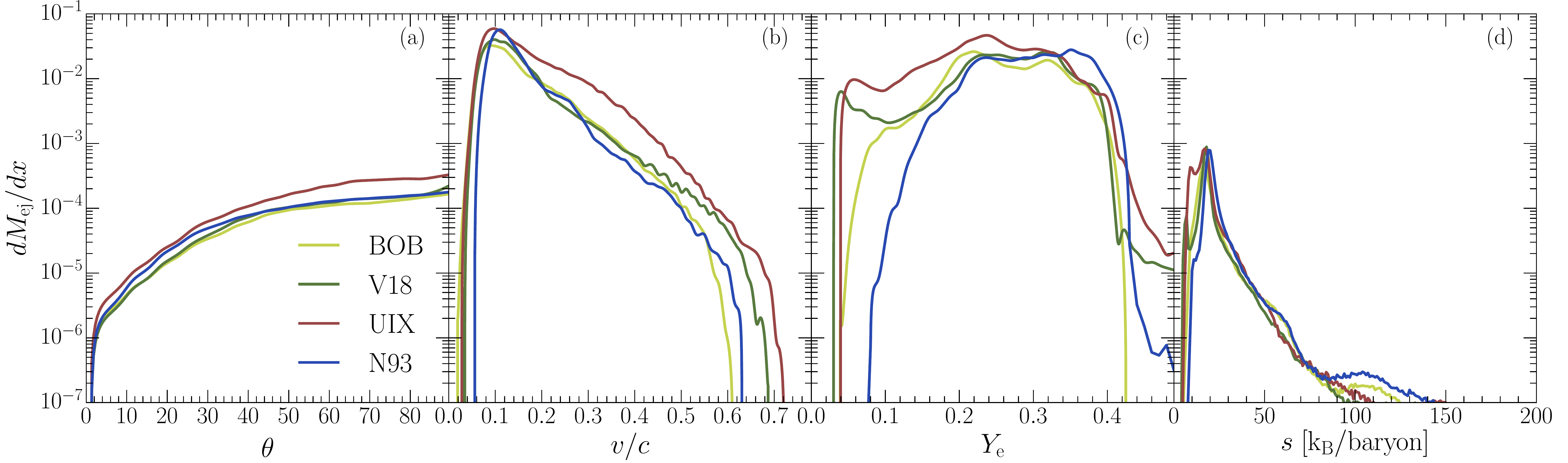}}
\vspace{-3mm}
\caption{
Distribution of the ejected mass as functions of the
polar angle $\theta$ (a),
velocity ratio $v/c$ (b),
electron fraction $Y_e$ (c), and
specific entropy $s$ (d). Curves are normalized with respect to the total ejected mass values $M_{\rm ej}$ enlisted in Tab. \ref{t:sim}.
}
\label{f:ej}
\end{figure*}

In the following we analyze in more detail properties of the remnant
that is formed after the merger.
Fig.~\ref{f:mb} (upper panel)
shows the profiles of the enclosed baryonic mass $M_B$
as a function of the spherical radius $r$ at $t = 15\,$ms after the merger;
we perform such calculation for all the cases we have investigated
by computing volume integrals of the conserved rest-mass density $D$
up to each different spherical radius $r$.
The results show that the softest of our EOS sample, UIX,
leads to the most compact remnant and viceversa for the BOB EOS,
which is the stiffest EOS.
For every case, a thick dashed vertical line denotes the $M_\text{obj}$ position,
according to Table~\ref{t:sim}
(although due to the density-cutoff procedure $M_\text{obj}$
is not the mass of a spherical object).
Again, the V18 and N93 EOSs lead to very similar profiles,
as also for other global properties in Table~\ref{t:eos}.
Thin dashed vertical lines indicate the radii of the $T=50\mev$
Kepler configurations (Table~\ref{t:eos}) for a qualitative comparison.
In fact we note for BOB,V18,N93 quite similar values and trends
as the remnant radii,
whereas the UIX remnant radius at $t=15\;$ms is significantly smaller
than the one of stable Kepler rotation,
which points again to the commencing collapse of this object.

In Fig.~\ref{f:mb} (lower panel)
we illustrate the angular velocity profiles for the four remnants at $t=15\;$ms.
In particular,
we perform averages in the azimuthal direction on the equatorial plane ($z=0$)
and over a time window of $1\,$ms at $t=15\,$ms,
see Eq.~(\ref{e:oav}),
so as to obtain functions that depend only on the cylindrical radius $r$
from the center of the grid.
The EOSs considered here exhibit a maximum of the averaged angular velocity
located at approximately $r\approx8\;$km,
roughly corresponding to the position where the two hot spots appeared
at the beginning of the merger \cite{Hanauske2016}.
Unsurprisingly, the maximum reached value is highest for the UIX EOS,
a feature which is fully compatible with our findings in the PSD distribution,
and also seen in Fig.~\ref{f:maxrhot}(d).
Also in this panel the horizontal dashed lines denote the Kepler frequencies
of the $T=50\mev$ $\mmax$ configurations for a qualitative comparison.
Once again the unstable nature of the UIX simulation
is confirmed by local rotation substantially above Kepler frequency.
We also point out that the profiles we have determined are in agreement
with results related to other EOSs
(see, e.g., \cite{Hanauske2016,DePietri2020})
and that, according to our results with the V18 EOS in Ref.~\cite{Figura2020},
we expect that such average profiles remain robust even when the same simulation
is carried out with the hybrid EOS approach,
choosing reasonable values for the $\Gamma_\text{th}$ parameter.

Finally,
we study in detail the properties of the dynamical ejecta in Fig.~\ref{f:ej};
in particular,
we characterize the mass ejection dependence on the polar angle
[panel (a); here, $0^\circ$ refers to the $z$ axis,
while $90^\circ$ is representative of the equatorial plane],
the velocity ratio $v/c$ (b),
the electron fraction $Y_e$ (c),
and the specific entropy $s$ (d).
In particular, the specific entropy
(which is directly related to the temperature of the ejected matter),
the velocity, and the electron fraction represent the most important quantities
in order to characterize the r-process nucleosynthesis in the outflows
\cite{Hotokezaka2015MNRAS}.

We determine that the emission increases almost monotonically
with the polar angle for all our cases,
with this behavior being particularly evident up to about $50^\circ$,
while the curves flatten out after this angle.
Thus the tidal disruption of matter generates ejecta
which are mostly distributed close to the equatorial plane
(labeled as equatorial ejecta in \cite{Vincent2020}).

Also the velocity distribution
shows common features for all the cases we analyze:
in particular, a clear peak at $v/c \approx 0.1$ is present,
after which the distribution decreases up to values between $0.6$ and $0.75$.
The tails of the distribution are ordered according to the stiffness of the EOS,
i.e., softer EOSs (UIX) eject more energetic matter.
These features are compatible with the cases analyzed
in Ref.~\cite{Bernuzzi2020b},
where different EOSs were considered.

The electron fraction represents a crucial parameter in order to determine
which elements can be created by the r-process;
indeed, heavy elements ($A \gtrsim 120$) are created via neutron-rich ejecta
($Y_e \lesssim 0.25$) \cite{Lippuner2015,Vincent2020},
whereas neutron-poor ejecta produce elements with lower masses.
Different $Y_e$ distributions also have an impact on the kilonova signal:
neutron-rich ejecta favor the so-called ``red'' kilonovae,
peaking in the infrared, while neutron-poor ejecta produce ``blue'' kilonovae
\cite{Metzger2017}.
Since our sample of simulations comprises only equal-mass mergers,
the resulting distributions favor the presence of the shocked component
over the tidal one \cite{Bernuzzi2020b}.
For our set of EOSs, in particular,
the ejecta distributions start at about $Y_e = 0.04$
with UIX and V18 showing a local maximum at that point.
A notable exception is the N93,
whose distribution is from $Y_e = 0.08$.
We note that N93 is the model with the largest nuclear symmetry energy
\cite{Wei2019,Wei2020}
and would thus inhibit more the ejection of neutron-rich matter.
For all our cases, the distributions cover a wide range of $Y_e$,
a feature which is however strongly dependent on the neutrino treatment
in the simulations,
which redistributes the electron number due to weak interactions,
see, e.g., Refs.~\cite{Radice2016,Radice2018a} for further details.

The study of the specific entropy, which,
as pointed out in Ref.~\cite{Papenfort2018},
has a close connection with the shock-heated matter in the ejecta,
shows that for all our EOSs,
a major fraction of the ejecta is characterized by low temperatures.
Interestingly, all our cases show a prominent peak corresponding to
$s \sim 20\;k_B/{\rm baryon}$
followed by a continuous drop.
UIX indeed shows a lower satellite peak at about
$s \sim 10\;k_B/{\rm baryon}$,
i.e., a higher fraction of matter is expelled at lower temperature
for this unstable transient state.

We close by repeating that without good theoretical and numerical control
on neutrino radiation and nuclear viscosity
no reliable quantitative predictions for
(dynamical and secular) ejecta properties can be made currently.

\section{Summary}
\label{s:end}

We have presented the first simulations of a $2\times1.35\ms$
NS merger employing four microscopic temperature-dependent EOSs
derived in the BHF formalism
that fulfil all current empirical constraints by nuclear phenomenology,
and also respect recent limits on maximum NS mass and deformability.
All simulations have been performed with a consistent treatment
of finite-temperature effects,
going beyond what is usually done in the hybrid EOS approach.
A detailed comparison of both approaches was made in Ref.~\cite{Figura2020}.
We presented in particular a detailed study of the GW and hydrodynamical properties
and focused on analyzing the mass distribution of the
post-merger remnant and the properties of the ejected matter.
We also examined the validity of several universal relations for these EOSs.

We found that two of the EOSs, V18 and N93
(with $\mmax\approx2.3\!-\!2.4\ms)$,
are good candidates for a realistic modeling of the GW170817 event,
whereas the UIX is too soft with a low $\mmax\approx2.0\ms$
and related immediate onset of instability of the remnant.
The BOB EOS on the other hand is rather stiff
($\mmax\approx2.5\ms)$,
and the only one that would be able
to support even a static object as massive as the GW170817 remnant,
which makes it also appear less realistic.

However, in order to draw firm conclusions of this kind,
still a lot of progress in the theoretical and numerical modeling
of the post-merger phase is required,
in particular an accurate quantitative understanding of
viscosity, heat transfer, neutrino reactions, magnetic fields, etc.,
in the relevant superdense matter.
This is absolutely essential for a realistic modeling of the
temporal evolution of the remnant and the related GW signal.
Only then can really quantitative constraints on the EOS be deduced
from future observations of the postmerger GW signal.

\section*{Acknowledgments}

We acknowledge useful discussions with
R.~De~Pietri, D.~Radice, L.~Rezzolla, and K.~Takami.
Partial support comes from ``PHAROS,'' COST Action CA16214.
Simulations have been carried out on the
\hbox{MARCONI} cluster at CINECA, Italy.
This work is also sponsored by
the National Natural Science Foundation of China under Grant
Nos.~11475045, 11975077
and the China Scholarship Council, No.~201806100066.

\newcommand{\nphysa}{Nuclear Physics A}
\bibliographystyle{apsrev4-1-noeprint}
\bibliography{aeireferences} 

\end{document}